\documentclass[sigconf]{acmart}

\usepackage{amsmath}
\usepackage{booktabs}
\usepackage{float}
\usepackage{multirow}
\usepackage{pifont}
\usepackage{tabularx}
\usepackage{xspace}
\setcopyright{acmlicensed}
\acmConference[MM '26]{The 34th ACM International Conference on Multimedia}{November 10--14, 2026}{Rio de Janeiro, Brazil}
\acmBooktitle{Proceedings of the 34th ACM International Conference on Multimedia (MM '26), November 10--14, 2026}
\acmYear{2026}
\copyrightyear{2026}
\acmISBN{}
\acmDOI{}

\newcommand{\rasg}{\textsc{RASG}\xspace}
\newcommand{\rvt}{\textsc{RVT}\xspace}
\newcommand{\sbr}{\textsc{SBR}\xspace}
\newcommand{\waf}{\ensuremath{\mathrm{WAF}}\xspace}
\newcommand{\cmark}{\ding{51}}
\newcommand{\xmark}{\ding{55}}

\begin{document}

\title[From Expression to Reaction]{From Expression to Reaction: Role-aware Visual Transfer and Stimulus-guided Reasoning for Interlocutor Emotion Recognition}
\author{Wei Wang}
\affiliation{\institution{Guangdong University of Technology}\city{Guangzhou}\country{China}}
\email{ww9362028@gmail.com}
\author{Zhaowu Li}
\affiliation{\institution{Guangdong University of Technology}\city{Guangzhou}\country{China}}
\email{lizhaowu1@mails.gdut.edu.cn}
\author{Jianjie Luo}
\affiliation{\institution{Guangdong University of Technology}\city{Guangzhou}\country{China}}
\email{jianjieluo@gdut.edu.cn}
\author{Fu Lee Wang}
\affiliation{\institution{Hong Kong Metropolitan University}\city{Hong Kong}\country{China}}
\email{pwang@hkmu.edu.hk}
\author{Lap-Kei Lee}
\affiliation{\institution{Hong Kong Metropolitan University}\city{Hong Kong}\country{China}}
\email{lklee@hkmu.edu.hk}
\author{Zhenguo Yang}
\authornote{Corresponding author.}
\affiliation{\institution{Guangdong University of Technology}\city{Guangzhou}\country{China}}
\email{yzg@gdut.edu.cn}
\renewcommand{\shortauthors}{Wei Wang et al.}

\begin{abstract}
In this paper, we propose a Role-aware Stimulus-guided (RASG) framework for interlocutor emotion recognition, which predicts listener emotions from listener-only videos and speaker-only audios. RASG consists of Role-aware Visual Transfer (RVT) and Stimulus-guided Boundary Reasoning (SBR) modules, which address supervision mismatch due to the lack of labeled listener data and ambiguity among visually similar listener reactions whose interpretation depends on speaker context, respectively. More specifically, RVT selects speaker samples whose facial expressions support their emotion labels. It then filters listener tracks and uses reliable pseudo-labels to train a listener-centric visual expert. SBR uses a two-class language reasoner only when the visual model is uncertain. It treats speaker audio and text as context rather than direct emotion evidence to distinguish similar listener reactions. Experiments conducted on MER-Cross dataset shows that RASG achieves 76.25\% on MER-Cross and improves the performance of the baseline over 17\%. Our team ranks second in Track 1 (MER-Cross) of the MER Grand Challenge at ACM MM 2026.
\end{abstract}



\maketitle

\section{Introduction}
Modeling semantic relationships between visual content and natural language has been widely explored in multimodal learning \cite{luo2023semantic,luo2025exploring,luo2024unleashing,luo2021cocobert,pan2022autocaptions}. In conversational settings, multimodal understanding also involves interpreting the emotions expressed by speakers and the reactions of listeners. Speakers convey affect through words, voice, and facial behavior; listeners show how they receive and respond to these utterances. Interlocutor emotion recognition aims to infer listener emotion from listener video, speaker audio, and the corresponding transcript \cite{lian2026mer}. The capability supports affect-aware tutoring systems \cite{gordon2016affective} and socially interactive robots \cite{ceha2019curiosity}, which must adapt to users' emotional responses. The MER series has explored robustness, semi-supervised learning, and large-language-model-based affective understanding \cite{lian2023mer,lian2025mer}, while MER-Cross extends the benchmark from individual speakers to interlocutors \cite{lian2026mer}. MER-Cross defines six listener emotions: neutral, sad, worried, happy, surprise, and angry \cite{lian2026mer}.

\begin{figure*}[t]
  \centering
  \includegraphics[width=\textwidth]{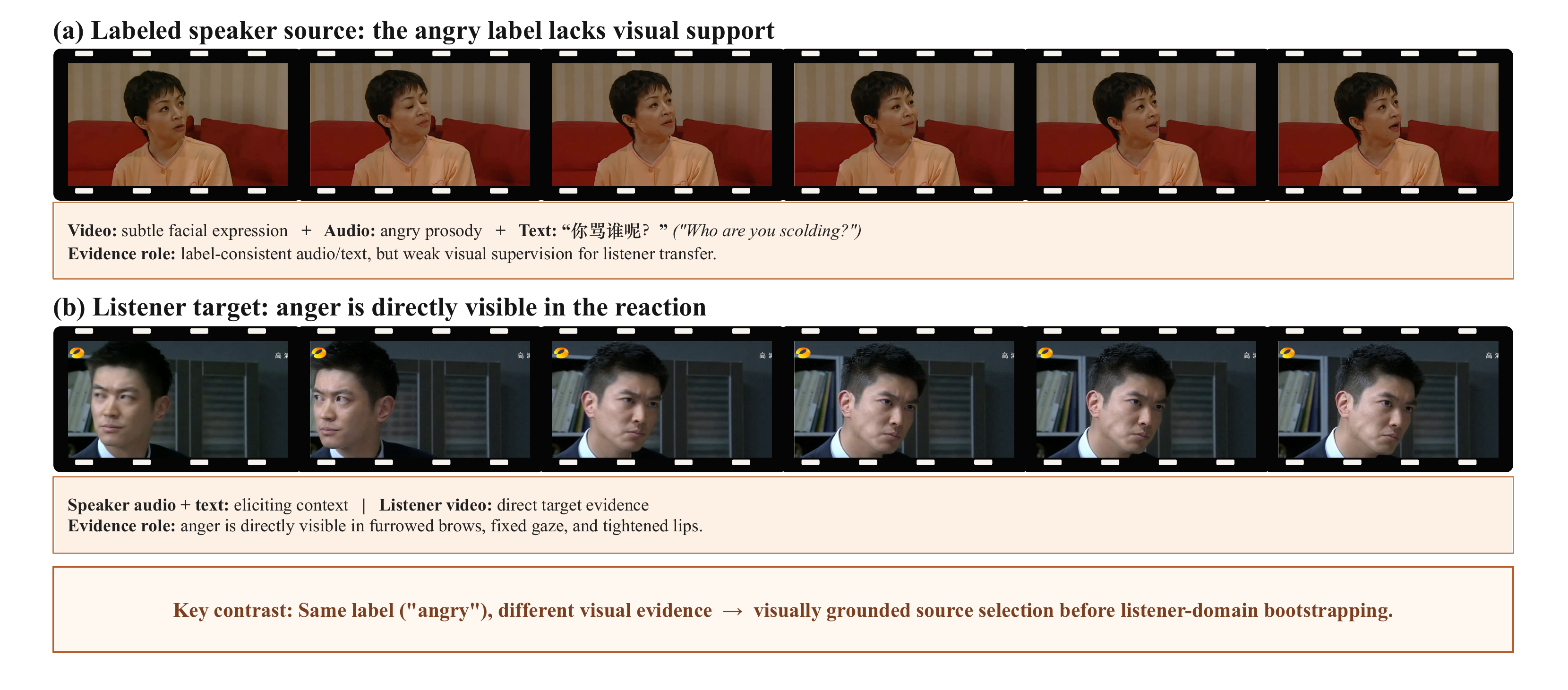}
  \caption{Role-conditioned evidence mismatch. (a) The speaker's angry label is supported mainly by audio and text despite weak facial affect. (b) Listener anger is directly visible; speaker audio and text provide only the eliciting context.}
  \Description{Two filmstrips compare an angry speaker sample with an angry listener sample. The speaker's facial affect remains subtle even though the speech is angry, whereas the listener directly manifests anger through facial behavior.}
  \label{fig:motivation}
\end{figure*}

Interlocutor emotion recognition poses three challenges: role mismatch, unreliable visual supervision, and distinct roles for listener video and speaker context. First, training labels describe speaker emotions, while predictions target listeners without labeled training examples \cite{lian2026mer}. Second, speaker emotion labels may reflect prosody or lexical content despite a neutral face, making them unreliable supervision for a listener-visual model. Third, speaker audio and text describe what may have elicited a reaction rather than independently observing the listener's state. Treating all three modalities as evidence of one person may therefore transfer speaker affect to listener predictions. Speaker-centered conversation models \cite{majumder2019dialoguernn,hu2021mmgcn,lian2023gcnet} capture temporal and relational context, yet their target emotion remains attached to the active speaker. Semi-supervised approaches \cite{sohn2020fixmatch,zhang2021flexmatch,chen2023classbalanced,li2023mining} can exploit unlabeled data, but errors in the initial model may be reinforced during pseudo-labeling. Language-model-based methods \cite{cheng2024emotionllama,lian2025affectgpt,huang2025affectivecot,jing2026causalerc} provide contextual emotion reasoning, although unconstrained predictions remain sensitive to prompts and spurious cues \cite{zhao2021calibrate,ludan2023explanation}. This highlights the importance of selecting and organizing relevant contextual information for multimodal reasoning \cite{liu2026structured}. The aforementioned limitations create a role-conditioned visual transfer gap between labeled speaker sources and unlabeled listener targets. As illustrated in Figure~\ref{fig:motivation}, even for the same emotion class, the speaker expressions used for training can differ substantially from the listener reactions encountered at inference time. Visual supervision from labeled speakers therefore cannot be transferred indiscriminately to listeners, motivating visually grounded source selection and listener-domain bootstrapping.

In this paper, we propose a listener-centric framework that learns a visual model without labeled listener data and uses speaker context only when the visual evidence is ambiguous. The framework contains two modules: Role-aware Visual Transfer and Stimulus-guided Boundary Reasoning. The first module retains speaker samples whose facial evidence agrees with their emotion labels and uses them to train an initial visual model. It then mines non-speaking listener tracks from an unlabeled conversational pool and accepts pseudo-labels under confidence, temporal, and cross-view consistency constraints. The second module activates only for low-margin visual predictions. It represents speaker audio and text as structured cues about the eliciting stimulus and restricts language-model reasoning to the two most plausible visual classes. The framework thus uses context without overriding confident listener-visual predictions.

Our contributions are fourfold:
\begin{itemize}
  \item We propose a listener-centric framework that combines direct listener-visual evidence with speaker stimulus information for interlocutor emotion recognition when no labeled listener data are available.
  \item We devise a Role-aware Visual Transfer module to filter visually inconsistent speaker labels and progressively build listener supervision from unlabeled non-speaking tracks.
  \item We develop a Stimulus-guided Boundary Reasoning module that applies pair-constrained language reasoning only to uncertain visual predictions and treats speaker content as eliciting context rather than target-emotion evidence.
  \item We conduct extensive experiments on MER-Cross, where \rasg achieves 76.25\% \waf and outperforms the official CLIP-large baseline by 17.37 percentage points.
\end{itemize}

\begin{figure*}[t]
  \centering
  \includegraphics[width=\textwidth]{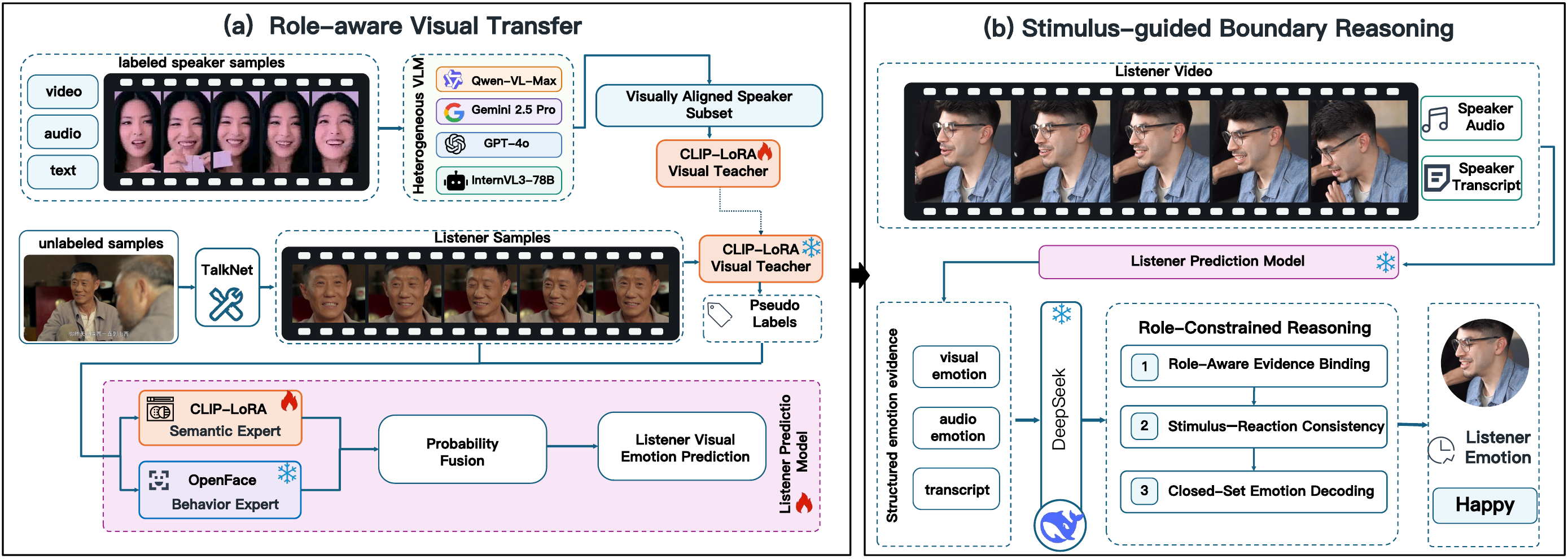}
  \caption{Overview of \rasg. (a) \rvt: Visually Aligned Speaker Supervision, Listener-domain Bootstrapping, and Listener Visual Expert Training. (b) \sbr: Low-margin Uncertainty Routing, Interaction Evidence Construction, and Top-2 Pair-constrained Reasoning.}
  \Description{The upper panel shows visually aligned speaker supervision, listener-domain bootstrapping, and listener visual expert training. The lower panel routes low-margin predictions, constructs role-separated interaction evidence, and uses a DeepSeek reasoner to select within the visual Top-2 candidate pair.}
  \label{fig:framework}
\end{figure*}

\section{Related Work}
\subsection{Conversational Emotion Recognition and Listener Modeling}
Early conversational emotion recognition methods modeled contextual and interpersonal dependencies while predicting the emotion of the current speaker. CMN stored multimodal histories and retrieved inter-speaker information through attention \cite{hazarika2018cmn}, while ICON encoded self- and inter-speaker influences in shared conversational memory \cite{hazarika2018icon}. DialogueRNN tracked global context, party states, and emotion states \cite{majumder2019dialoguernn}, and DialogueGCN propagated information over intra- and inter-speaker graph edges \cite{ghosal2019dialoguegcn}. COSMIC further incorporated commonsense knowledge about mental states, events, and causal relations \cite{ghosal2020cosmic}, whereas MMGCN modeled cross-modal and speaker dependencies with graph convolutions \cite{hu2021mmgcn}. Despite their different context-modeling mechanisms, these methods assume that the utterance and its modalities describe the labeled speaker rather than a listener's reaction.
Listener-centered research has primarily focused on reaction synthesis. Ng et al. \cite{ng2022learning} generated multiple plausible listener motions from speaker audio and facial motion, while Song et al. \cite{song2023listener} modeled listener motion with discrete codewords conditioned on emotional context. Both require paired speaker-listener data to generate reactions rather than recognize emotions in authentic listener videos.

\subsection{Visual Transfer with Reliable Pseudo-supervision}
Pseudo-labeling assigns model predictions to unlabeled samples as training targets \cite{lee2013pseudo}, but erroneous predictions may be reinforced. FixMatch improved reliability through confidence-based weak-to-strong consistency \cite{sohn2020fixmatch}, while class-balanced pseudo-labeling mitigated label imbalance in multimodal emotion recognition \cite{chen2023classbalanced}. However, these methods assume consistent target semantics and do not verify whether an emotion is visually observable or belongs to the correct conversational role.
Li et al. \cite{li2023mining} selected high-quality emotion samples using model agreement and majority voting, which may retain shared systematic biases. TalkNet detected active speakers through temporal and cross-modal modeling \cite{tao2021talknet}, but speaking status alone cannot determine whether a non-speaking track provides a valid listener reaction or reliable emotion supervision.

\subsection{Contextual Emotion Reasoning with Large Language Models}
Recent studies have used large language models for contextual emotion reasoning. Emotion-LLaMA aligned multimodal encoders with a language model through instruction tuning \cite{cheng2024emotionllama}, while AffectGPT combined multimodal pre-fusion, descriptive annotations, and unified emotion evaluation \cite{lian2025affectgpt}. Affective-CoT further decomposed multimodal emotion reasoning into a hierarchical cognitive process \cite{huang2025affectivecot}. These methods generally reason over the full emotion label space under the assumption that visual, acoustic, and textual evidence all describe the same emotional subject.
CausalERC introduced causal prompting for contextual emotion recognition \cite{jing2026causalerc}, but still predicted the current speaker's emotion. Zhao et al. \cite{zhao2021calibrate} corrected prompt-induced label biases using content-free inputs, yet such general calibration does not impose the role-specific constraints needed to distinguish direct listener evidence from the speaker context that elicits the listener's reaction.

\section{Methodology}
\subsection{Problem Definition}
Interlocutor emotion recognition aims to infer a listener's emotion from the listener video and the corresponding speaker audio and transcript. Generally, let
\begin{equation}
\mathcal{D}_{s}=\left\{\left(V_{i}^{s},A_{i}^{s},T_{i}^{s},y_{i}^{s}\right)\right\}_{i=1}^{N_{s}}
\label{eq:speakerdata}
\end{equation}
denote the labeled speaker dataset, where $V_{i}^{s}$, $A_{i}^{s}$, and $T_{i}^{s}$ represent the video, audio, and transcript, respectively; $y_{i}^{s}$ is the speaker emotion label; and $N_{s}$ is the number of labeled speaker samples.

An unlabeled conversational pool is also available:
\begin{equation}
\mathcal{D}_{u}=\left\{\left(V_{j}^{u},A_{j}^{u},T_{j}^{u}\right)\right\}_{j=1}^{N_{u}},
\label{eq:unlabeleddata}
\end{equation}
where $N_{u}$ is the number of unlabeled conversational samples. Their conversational roles and emotion labels are unknown.

At inference, an instance contains listener video $V^{l}$ and the corresponding speaker audio $A^{s}$ and transcript $T^{s}$. The objective is
\begin{equation}
\widehat{y}^{l}=F\left(V^{l},A^{s},T^{s}\right),\qquad \widehat{y}^{l}\in\mathcal{C},
\label{eq:objective}
\end{equation}
where $\mathcal{C}=\{\text{neutral},\text{angry},\text{happy},\text{sad},\text{worried},\text{surprise}\}$. Unlike conventional multimodal emotion recognition, the modalities describe two participants: video directly captures the listener's reaction, whereas audio and text characterize the speaker's preceding stimulus. Role separation determines how each modality is used.

\subsection{Overview of the Framework}
Figure~\ref{fig:framework} presents the \textbf{Role-aware Stimulus-guided (\rasg) framework} for two challenges. First, no labeled listener samples are available; existing speaker labels may reflect audio, text, or multimodal evidence and thus provide unreliable facial supervision. Second, subtle listener expressions can look similar; speaker context helps disambiguate them but does not directly indicate listener emotion. \textbf{Role-aware Visual Transfer (\rvt)} selects visually reliable speaker supervision and progressively transfers it to listeners. \textbf{Stimulus-guided Boundary Reasoning (\sbr)} uses speaker audio and text as the eliciting stimulus to resolve ambiguous visual predictions. Thus, \rvt builds a listener-oriented visual model, and \sbr resolves context-dependent boundaries without overriding confident visual evidence.

\subsection{Role-aware Visual Transfer}
Speaker and listener labels follow different evidence structures. A speaker may express anger prosodically while keeping a neutral face; the label is valid for multimodal speaker recognition but noisy for facial learning. Moreover, the unlabeled pool mixes active speakers and non-speaking listeners. Direct source training thus causes label-evidence mismatch, while pseudo-labeling may assign supervision to the wrong role. \rvt addresses both problems by selecting speaker labels supported by facial behavior, training an initial visual teacher, and using it to annotate strictly filtered listener tracks for listener-domain adaptation.

\subsubsection{Visually Aligned Speaker Supervision}
We use a heterogeneous committee of $M$ vision-language models: Qwen-VL-Max \cite{bai2023qwenvl}, GPT-4o \cite{openai2024gpt4o}, Gemini 2.5 Pro \cite{gemini2025report}, and InternVL3-78B \cite{zhu2025internvl3}. Each independently judges the speaker's visual emotion. Let $\widetilde{y}_{i,m}^{v}$ be the $m$-th VLM prediction for $V_{i}^{s}$. Their agreement is
\begin{equation}
q_{i}=\frac{1}{M}\sum_{m=1}^{M}\mathbb{I}\left(\widetilde{y}_{i,m}^{v}=y_{i}^{s}\right),
\label{eq:agreement}
\end{equation}
where $\mathbb{I}(\cdot)$ is the indicator. The committee tests whether the original dataset label is recoverable from visual evidence alone without replacing it. We retain
\begin{equation}
\mathcal{D}_{s}^{va}=\left\{\left(V_{i}^{s},y_{i}^{s}\right)\mid q_{i}\geq\tau_{c}\right\},
\label{eq:alignedsource}
\end{equation}
where $\tau_{c}$ is the cross-model consensus threshold. The retained speaker labels are inferable without audio or text and thus better match listener-visual supervision \cite{li2023mining}.

A CLIP \cite{radford2021clip} encoder adapted on $\mathcal{D}_{s}^{va}$ with LoRA \cite{hu2022lora} yields the initial teacher $f_{T}$. The adaptation reduces the influence of speaker labels dominated by linguistic or acoustic evidence.

\subsubsection{Listener-domain Bootstrapping}
TalkNet \cite{tao2021talknet} detects speaking activity for each visible person in the 124,802 unlabeled videos. Face tracks remaining non-speaking throughout the corresponding speaker utterance are retained as strict listener samples.

Let $r_{j}\in\{0,1\}$ indicate whether an extracted face track satisfies the listener constraint. For each retained listener video $V_{j}^{l}$, the visual teacher produces
\begin{equation}
\mathbf{p}_{j}^{T}=\operatorname{softmax}\left(f_{T}\left(V_{j}^{l}\right)\right),
\label{eq:teacherprediction}
\end{equation}
where $\mathbf{p}_{j}^{T}\in\mathbb{R}^{|\mathcal{C}|}$ is the six-class distribution. Its pseudo-label and confidence are
\begin{equation}
\widehat{y}_{j}^{l}=\arg\max_{k\in\mathcal{C}}p_{j,k}^{T},\qquad c_{j}=\max_{k\in\mathcal{C}}p_{j,k}^{T}.
\label{eq:pseudolabelconfidence}
\end{equation}
We retain confident listener predictions:
\begin{equation}
\mathcal{D}_{l}^{p}=\left\{\left(V_{j}^{l},\widehat{y}_{j}^{l},c_{j}\right)\mid r_{j}=1,\;c_{j}\geq\tau_{p}\right\},
\label{eq:listenerpseudodata}
\end{equation}
where $\tau_{p}$ is the confidence threshold. Role filtering selects listener tracks, while confidence filtering ensures reliable visual emotion; together they create listener-domain supervision \cite{sohn2020fixmatch,chen2023classbalanced,li2023mining}.

\subsubsection{Listener Visual Expert Training}
Figure~\ref{fig:framework}(a) shows listener-domain CLIP adaptation \cite{radford2021clip}. Starting from the speaker-trained teacher, we freeze original CLIP parameters and insert trainable low-rank matrices. For projection $\mathbf{W}$, LoRA \cite{hu2022lora} gives
\begin{equation}
\mathbf{W}'=\mathbf{W}+\frac{\alpha}{r}\mathbf{B}\mathbf{A},
\label{eq:lora}
\end{equation}
where $\mathbf{A}\in\mathbb{R}^{r\times d}$ and $\mathbf{B}\in\mathbb{R}^{d'\times r}$ are trainable low-rank matrices, $r$ is the LoRA rank, and $\alpha$ controls update magnitude. Only $\mathbf{A}$, $\mathbf{B}$, and the emotion head are updated.

For listener video $V_{j}^{l}=\{x_{j,t}\}_{t=1}^{T_{j}}$, frame representations are
\begin{equation}
\mathbf{z}_{j,t}=E_{\mathrm{CLIP}}\left(x_{j,t};\mathbf{W}'\right).
\label{eq:clipframe}
\end{equation}
They are aggregated at video level:
\begin{equation}
\bar{\mathbf{z}}_{j}=\operatorname{Pool}\left(\mathbf{z}_{j,1},\ldots,\mathbf{z}_{j,T_{j}}\right),\qquad
\mathbf{p}_{j}^{c}=\operatorname{softmax}\left(\mathbf{W}_{c}\bar{\mathbf{z}}_{j}\right).
\label{eq:clipprediction}
\end{equation}
Here, $\operatorname{Pool}(\cdot)$ aggregates temporal features and $\mathbf{W}_{c}$ is the emotion head. Confidence-weighted adaptation downweights uncertain listener pseudo-labels, shifting the model toward listener facial reactions while retaining visually grounded speaker knowledge.

CLIP-LoRA captures holistic facial appearance and expression semantics, while subtle listener reactions may involve local facial actions, gaze changes, and head motion. We therefore add a complementary OpenFace behavior branch \cite{baltrusaitis2018openface}.

Let $O(V_{j}^{l})$ contain OpenFace action units, gaze, pose, and temporal descriptors for $V_{j}^{l}$. The branch predicts
\begin{equation}
\mathbf{p}_{j}^{b}=\operatorname{softmax}\left(f_{b}\left(O(V_{j}^{l})\right)\right),
\label{eq:openfaceprediction}
\end{equation}
where $f_{b}$ is the behavior classifier. The two expert predictions are combined at the probability level:
\begin{equation}
\mathbf{p}_{j}^{v}=\lambda\mathbf{p}_{j}^{c}+(1-\lambda)\mathbf{p}_{j}^{b},\qquad
\widehat{y}_{j}^{v}=\arg\max_{k\in\mathcal{C}}p_{j,k}^{v},
\label{eq:visualfusion}
\end{equation}
where $\lambda$ balances both branches. Using $\mathbf{p}_{j}^{v}$ rather than a hard label preserves the confidence and competing candidates for reasoning.

\subsection{Stimulus-guided Boundary Reasoning}
The visual expert handles prototypical reactions, but difficult samples often lie near class boundaries rather than simply exhibiting weak facial signals: subdued faces may indicate neutral, sad, or worried depending on the preceding interaction. Speaker audio and text describe the eliciting stimulus rather than the target person and cannot serve as direct listener-emotion evidence. \sbr therefore uses context only for uncertain visual predictions and restricts reasoning to the two most plausible visual classes. This can resolve specific visual ambiguity without replacing the visual model with unrestricted language-model inference \cite{cheng2024emotionllama,lian2025affectgpt,huang2025affectivecot}.

\subsubsection{Low-margin Uncertainty Routing}
Let $c_{(1)}$ and $c_{(2)}$ be the top two classes in $\mathbf{p}^{v}$, with probabilities $p_{(1)}^{v}$ and $p_{(2)}^{v}$. Their decision margin is
\begin{equation}
\Delta^{v}=p_{(1)}^{v}-p_{(2)}^{v}.
\label{eq:margin}
\end{equation}
A large margin indicates a clear visual preference requiring no contextual interpretation, whereas a small margin signals a boundary case \cite{geifman2019selective}. We define
\begin{equation}
\mathcal{C}_{2}^{v}=\left\{c_{(1)},c_{(2)}\right\},
\label{eq:top2pair}
\end{equation}
and activate the DeepSeek \cite{deepseek2024v3} reasoner only when
\begin{equation}
\Delta^{v}<\tau_{m},
\label{eq:routing}
\end{equation}
where $\tau_{m}$ is the routing threshold.

\subsubsection{Interaction Evidence Construction}
For routed samples, the official acoustic model predicts emotion $\widehat{y}^{a}$ and confidence $c^{a}$ from speaker audio $A^{s}$. These are combined with the complete listener distribution and corresponding transcript:
\begin{equation}
\mathcal{E}=\operatorname{Enc}\left(\text{listener},\mathbf{p}^{v},\Delta^{v},\mathcal{C}_{2}^{v},\widehat{y}^{a},c^{a},T^{s}\right).
\label{eq:evidence}
\end{equation}
This evidence contains the listener-visual distribution, decision margin, Top-2 candidates, speaker-audio prediction and confidence, and transcript. Role labels separate listener-target evidence from speaker-stimulus context.

\subsubsection{Top-2 Pair-constrained Reasoning}
The DeepSeek prompt performs three operations: \textbf{role-aware evidence binding} separates direct listener evidence from speaker stimulus evidence; \textbf{stimulus-reaction consistency reasoning} checks how speaker tone and content may elicit the reaction; and \textbf{pair-constrained emotion decoding} selects from $\mathcal{C}_{2}^{v}$.

For a low-margin sample, the reasoning result is
\begin{equation}
\widehat{y}^{r}=R_{\mathrm{DS}}\left(\mathcal{E},\mathcal{C}_{2}^{v};\mathcal{P}_{\mathrm{bind}},\mathcal{P}_{\mathrm{react}},\mathcal{P}_{\mathrm{pair}}\right),\quad
\widehat{y}^{r}\in\mathcal{C}_{2}^{v}.
\label{eq:reasoning}
\end{equation}
The final prediction is
\begin{equation}
\widehat{y}^{l}=\begin{cases}
c_{(1)}, & \Delta^{v}\geq\tau_{m},\\
\widehat{y}^{r}, & \Delta^{v}<\tau_{m}.
\end{cases}
\label{eq:finalprediction}
\end{equation}
This formulation separates responsibilities: the visual expert handles distinctive reactions, while DeepSeek resolves ambiguous boundaries. Restricting outputs to the visual Top-2 candidates prevents unconstrained labels from unrelated contextual cues.

These constraints also govern inference. The listener visual expert first produces the six-class distribution and decision margin, so listener evidence alone determines whether contextual reasoning is needed. Speaker audio and transcript are introduced only for samples routed as ambiguous. Unrouted samples retain the visual Top-1 class; routed samples are restricted to the two visually supported candidates. This ordering prevents speaker emotion from becoming the prediction target while allowing the preceding stimulus to resolve visually ambiguous boundaries.

\subsection{Training Objectives}
\rasg adds no task-specific loss: all trainable visual components use standard cross-entropy objectives, while the inference-only DeepSeek reasoner introduces no training loss.

We first train the visual teacher on aligned speaker samples:
\begin{equation}
\mathcal{L}_{T}=-\frac{1}{|\mathcal{D}_{s}^{va}|}\sum_{(V_{i}^{s},y_{i}^{s})\in\mathcal{D}_{s}^{va}}\log p_{i,y_{i}^{s}}^{T}.
\label{eq:teacherloss}
\end{equation}
The listener-domain CLIP branch then uses confidence-weighted pseudo-label supervision \cite{sohn2020fixmatch,chen2023classbalanced}:
\begin{equation}
\mathcal{L}_{c}=-\frac{1}{Z}\sum_{(V_{j}^{l},\widehat{y}_{j}^{l},c_{j})\in\mathcal{D}_{l}^{p}}c_{j}\log p_{j,\widehat{y}_{j}^{l}}^{c},
\label{eq:cliploss}
\end{equation}
where $Z=\sum_{j}c_{j}$ normalizes pseudo-label weights. The OpenFace behavior classifier uses the same listener supervision:
\begin{equation}
\mathcal{L}_{b}=-\frac{1}{Z}\sum_{(V_{j}^{l},\widehat{y}_{j}^{l},c_{j})\in\mathcal{D}_{l}^{p}}c_{j}\log p_{j,\widehat{y}_{j}^{l}}^{b}.
\label{eq:behaviorloss}
\end{equation}
The complete visual training objective is
\begin{equation}
\mathcal{L}_{\mathrm{RVT}}=\mathcal{L}_{T}+\lambda_{c}\mathcal{L}_{c}+\lambda_{b}\mathcal{L}_{b},
\label{eq:rvtloss}
\end{equation}
where $\lambda_{c}$ and $\lambda_{b}$ weight listener-domain CLIP adaptation and behavior training. Although joined in Eq.~\ref{eq:rvtloss}, the objectives are optimized sequentially as in Fig.~\ref{fig:framework}(a).

\section{Experiments}
\subsection{Datasets}
\hspace*{\parindent}\textbf{MER-Cross \cite{lian2026mer}.} The MER2026 interlocutor emotion recognition track provides 9,395 labeled speaker samples for training and 574 interlocutor samples for testing. Training samples contain speaker video, audio, transcript, and emotion labels; test samples contain listener video with speaker audio and transcript, targeting listener emotion. The six classes are neutral, angry, happy, sad, worried, and surprise. Speaker training labels and listener test labels create a role-conditioned supervision gap.

\textbf{MER-SEMI Unlabeled Pool \cite{lian2025mer}.} We use 124,802 unlabeled MER videos without listener annotations. Active-speaker detection retains non-speaking face tracks as candidate listener reactions. The pool serves only listener-domain bootstrapping; MER-Cross test labels are excluded from sample selection, pseudo-labeling, and visual model training.

\subsection{Baselines}
\hspace*{\parindent}\textbf{HuBERT \cite{hsu2021hubert}.} The official HuBERT-base baseline captures speaker audio cues rather than directly observing the listener.

\textbf{MacBERT \cite{cui2020macbert}.} MacBERT-large provides a lexical baseline by encoding speaker transcripts.

\textbf{VideoMAE \cite{tong2022videomae}.} The official VideoMAE-large baseline uses temporal video features to predict listener emotion.

\textbf{ResNet-FER2013 \cite{lian2026mer,he2016resnet}.} The FER2013-trained residual network provides a frame-level facial-expression baseline.

\textbf{EmoNet \cite{toisoul2021emonet}.} The facial affect baseline models categorical emotions, action units, and continuous affective dimensions.

\textbf{SENet-FER2013 \cite{lian2026mer,hu2018senet}.} Channel-wise recalibration strengthens the FER2013-trained convolutional baseline.

\textbf{CLIP \cite{radford2021clip}.} We report CLIP-base and CLIP-large, the strongest official unimodal baseline and our main comparison.

\textbf{Top1 Fusion \cite{lian2026mer}.} The baseline combines HuBERT-base audio, MacBERT-large text, and CLIP-large video for a shared prediction.

\textbf{Top2 Fusion \cite{lian2026mer}.} The baseline combines HuBERT-large/base, MacBERT-base/large, and CLIP-base/large to assess the benefit of greater multimodal feature diversity.

\textbf{OpenFace-CLIP Fusion \cite{radford2021clip,baltrusaitis2018openface}.} Our reference fuses holistic CLIP features with OpenFace action units, gaze, pose, and motion, without listener-domain adaptation or language reasoning.

\subsubsection{Evaluation Metrics}
We use weighted average F1-score (\waf) as the primary metric under the official protocol \cite{lian2026mer}:
\begin{equation}
\waf=\sum_{c\in\mathcal{C}}\frac{N_{c}}{N}\frac{2P_{c}R_{c}}{P_{c}+R_{c}},
\label{eq:waf}
\end{equation}
where $P_{c}$, $R_{c}$, and $N_{c}$ are class-$c$ precision, recall, and sample count, and $N$ is the total count. Accuracy (ACC) is used as the secondary metric. Because \waf is the official ranking metric and final challenge measure, the following tables focus on it.

\subsubsection{Implementation Details}
Listener faces are detected, aligned, and resized to $224\times224$ pixels. CLIP ViT-L/14 \cite{radford2021clip} is the main encoder and is adapted with LoRA \cite{hu2022lora}; OpenFace \cite{baltrusaitis2018openface} remains a fixed facial-behavior extractor; and TalkNet \cite{tao2021talknet} identifies non-speaking tracks in the unlabeled pool. All visual ablations use the same preprocessing and classification protocol.

The reasoner receives the full six-class visual distribution, Top-2 labels, decision margin, speaker audio prediction, and transcript. DeepSeek \cite{deepseek2024v3} is invoked only below the routing threshold and restricted to the visual Top-2 pair. All ablations use identical data and evaluation protocols, varying only the indicated component.

\subsection{Quantitative Performance of the Approaches}
\begin{table}[t]
\caption{Official MER-Cross baseline comparison \cite{lian2026mer} in \waf (\%). A: speaker audio; T: speaker text; V: listener video.}
\label{tab:main}
\centering
\small
\begin{tabular*}{\columnwidth}{@{\extracolsep{\fill}}lcc@{}}
\toprule
Method & Modality & \waf (\%) \\
\midrule
HuBERT-base \cite{hsu2021hubert} & A & 35.74 \\
MacBERT-large \cite{cui2020macbert} & T & 33.68 \\
VideoMAE-large \cite{tong2022videomae} & V & 41.08 \\
ResNet-FER2013 \cite{lian2026mer,he2016resnet} & V & 45.86 \\
EmoNet \cite{toisoul2021emonet} & V & 47.03 \\
SENet-FER2013 \cite{lian2026mer,hu2018senet} & V & 51.55 \\
CLIP-base \cite{radford2021clip} & V & 52.43 \\
CLIP-large \cite{radford2021clip} & V & 58.88 \\
Top1 Fusion \cite{lian2026mer} & A+T+V & 57.44 \\
Top2 Fusion \cite{lian2026mer} & A+T+V & 55.24 \\
\textbf{\rasg (Ours)} & A+T+V & \textbf{76.25} \\
\bottomrule
\end{tabular*}
\end{table}

\begin{table}[t]
\caption{Top-three teams in Track 1 (MER-Cross) of the MER Grand Challenge at ACM MM 2026. Our team is in bold.}
\label{tab:leaderboard}
\centering
\small
\begin{tabular*}{\columnwidth}{@{\extracolsep{\fill}}lcc@{}}
\toprule
Team & \waf (\%) $\uparrow$ & Rank \\
\midrule
HLT-HITSZ & 78.31 & 1 \\
\textbf{WislabGDUT (Ours)} & \textbf{76.25} & \textbf{2} \\
CASIA & 74.07 & 3 \\
\bottomrule
\end{tabular*}
\end{table}

Table~\ref{tab:main} compares \rasg with the official audio-, text-, video-, and multimodal baselines in \waf. Three findings follow. \textbf{1)} Listener video directly observes the target and is more informative than speaker audio or text: CLIP-large reaches 58.88\%, compared with 35.74\% for HuBERT-base and 33.68\% for MacBERT-large. \textbf{2)} Direct multimodal fusion is ineffective: Top1 and Top2 obtain 57.44\% and 55.24\%, respectively, both below CLIP-large. \textbf{3)} \rasg achieves 76.25\%, exceeding CLIP-large by 17.37 points and Top1 by 18.81 points, confirming the benefit of listener-oriented visual transfer and selective use of speaker context.

\textbf{MER-Cross Evaluation.} Table~\ref{tab:leaderboard} reports the top-three leaderboard results for Track 1 (MER-Cross) of the MER Grand Challenge at ACM MM 2026, which focuses on interlocutor emotion recognition. Our team, WislabGDUT, achieves a \waf of 76.25\% with the proposed \rasg framework and secures second place.

\subsection{Ablation Studies}
\begin{table}[t]
\caption{Ablation study of \rasg on MER-Cross.}
\label{tab:ablation}
\centering
\small
\setlength{\tabcolsep}{3pt}
\begin{tabular}{lccccc}
\toprule
Method & OpenFace & CLIP & \rvt & \sbr & \waf (\%) \\
\midrule
OpenFace + CLIP & \cmark & \cmark & \xmark & \xmark & 64.28 \\
OpenFace + fine-tuned CLIP & \cmark & \cmark & \cmark & \xmark & 73.42 \\
\textbf{Full \rasg} & \cmark & \cmark & \cmark & \cmark & \textbf{76.25} \\
\bottomrule
\end{tabular}
\end{table}

Table~\ref{tab:ablation} evaluates \rvt and \sbr by adding them progressively to the OpenFace + CLIP reference. Two findings follow. \textbf{1)} \rvt improves \waf from 64.28\% to 73.42\%, demonstrating the main benefit of role-aligned listener supervision. \textbf{2)} \sbr further raises \waf to 76.25\%, showing that speaker context provides complementary gains for low-margin cases. The full framework therefore improves the OpenFace + CLIP reference by 11.97 points.

\subsection{Effectiveness of \rvt and \sbr with the CLIP-large Baseline}
\begin{table}[t]
\caption{Component gains over the CLIP-large baseline.}
\label{tab:gains}
\centering
\small
\setlength{\tabcolsep}{3pt}
\begin{tabularx}{\columnwidth}{@{}Xrrr@{}}
\toprule
Stage & \waf (\%) & Increment & Cum. Gain \\
\midrule
Official CLIP-large & 58.88 & -- & -- \\
+ OpenFace behavior expert & 64.28 & +5.40 & +5.40 \\
+ Role-aware Visual Transfer & 73.42 & +9.14 & +14.54 \\
+ \sbr & \textbf{76.25} & +2.83 & \textbf{+17.37} \\
\bottomrule
\end{tabularx}
\end{table}

Table~\ref{tab:gains} decomposes the 17.37-point gain over the CLIP-large baseline. Three findings follow. \textbf{1)} OpenFace contributes 5.40 points through local facial behavior cues. \textbf{2)} \rvt provides the largest gain of 9.14 points, confirming the importance of listener-domain supervision. \textbf{3)} \sbr adds 2.83 points by resolving ambiguous low-margin cases without overriding confident visual predictions.

The decomposition also clarifies the modules' complementary roles. \rvt yields the larger gain by addressing role-conditioned supervision before listener prediction, while \sbr adds a smaller gain on the remaining ambiguous cases. Since \sbr operates only on low-margin samples within the visual Top-2 pair, it complements rather than replaces the listener-oriented visual expert.

\section{Conclusion}
\rasg addresses interlocutor emotion recognition without labeled listener data. \rvt filters visually consistent speaker samples and bootstraps pseudo-labeled listener tracks, while \sbr uses speaker context only for uncertain visual predictions. \rasg achieves 76.25\% \waf on MER-Cross, outperforming the official CLIP-large baseline by 17.37 percentage points. \rasg achieved second place in Track 1 (MER-Cross) of the MER Grand Challenge at ACM MM 2026, and our results show the complementary value of role-aware visual transfer and constrained stimulus reasoning.

\bibliographystyle{ACM-Reference-Format}
\bibliography{references}

\end{document}